\pdfoutput=1

\documentclass[journal,comsoc]{IEEEtran}

\usepackage[T1]{fontenc}
\usepackage[utf8]{inputenc}
\usepackage{cite}
\usepackage{amsmath,amssymb}
\usepackage{graphicx}
\graphicspath{{./}}
\usepackage{booktabs}
\usepackage{array}
\usepackage{textcomp}
\usepackage{xcolor}
\usepackage{eso-pic}
\usepackage{url}
\usepackage[hidelinks]{hyperref}
\hypersetup{
	pdftitle={Is Classical Error Control Ready for IMT-2030? Information-Loss Interfaces in Channel Estimation, FEC, CRC, and ARQ for Machine-Type Services},
	pdfauthor={Igor B. Palhano}
}

\begin{document}
	
	\title{Is Classical Error Control Ready for IMT-2030? Information-Loss Interfaces in Channel Estimation, FEC, CRC, and ARQ for Machine-Type Services}
	
	\author{Igor B. Palhano,~\IEEEmembership{Student Member,~IEEE}%
		\thanks{I.~B.~Palhano is with the Federal University of Ceará (UFC), Fortaleza, Brazil, and the École Nationale Supérieure d'Ingénieurs de Caen (ENSICAEN), Caen, France.}%
	}
	
	\markboth{Preprint}%
	{Palhano: Is Classical Error Control Ready for IMT-2030?}

	\maketitle

	\AddToShipoutPictureBG*{%
		\AtPageLowerLeft{%
			\raisebox{0.42in}{\makebox[\paperwidth]{\parbox{\textwidth}{\centering\scriptsize\color{black!75}
				This work has been submitted to the IEEE for possible publication. Copyright may be transferred without notice, after which this version may no longer be accessible.
			}}}%
		}%
	}
	
	\begin{abstract}
		Fault tolerance and structured error control in mobile communication networks---historically implemented as a modular cascade of pilot-aided channel estimation (CE), forward error correction (FEC), cyclic redundancy check (CRC), and hybrid automatic repeat request (HARQ)---remain foundational to 5G New Radio (NR). Although finite-blocklength coding, joint estimation-decoding, and grant-free access have been active academic topics for over a decade, commercial 3GPP specifications up to Release~18 still enforce rigid cross-layer boundaries between these blocks. In short-packet machine-type communications (MTC) for IMT-2030 (6G), inserting separate redundancy for channel training, error correction, and error detection shrinks the net payload entropy resolved per channel slot, while standardized UE processing timelines and HARQ round-trip delays collide with sub-millisecond reliability targets. Combining a multi-user MIMO short-packet simulation with real 3GPP TS~38.214 numerology latency budgets, this article diagnoses the structural bottlenecks of the legacy reliability pipeline and outlines an interoperable roadmap for 3GPP Release~20+---spanning soft-output universal decoding, sensing-assisted priors, grant-free joint detection, and task-oriented error control.
	\end{abstract}
	
	\section{Fault Tolerance Is Not Dead, But Its Modular Boundaries Are}
	\IEEEPARstart{O}{ver} the past three decades, physical- and link-layer reliability in cellular networks has been governed by an extraordinarily successful architectural abstraction. Following Shannon's separation principles and modular protocol layering, wireless transceivers decompose the problem of delivering trustworthy bits over noisy, time-varying channels into four sequential stages: pilot-aided channel estimation (CE) to acquire channel state information (CSI), forward error correction (FEC) to combat noise and fading through structured bit redundancy, cyclic redundancy check (CRC) to verify decoding integrity, and hybrid automatic repeat request (HARQ/ARQ) to coordinate retransmissions. Because modern 5G New Radio (NR) LDPC and polar codes operate close to theoretical limits for long payloads, industry attention in early 6G discussions has shifted heavily toward generative AI, reconfigurable intelligent surfaces, and sub-THz bands, often treating structured error control as a mature, largely settled layer.
	
	Yet finite-blocklength (FBL) information theory and short-packet reliability have remained vibrant academic fields precisely because the asymptotic assumptions of modular error control break down for short machine-type bursts~\cite{durisi2016,polyanskiy2010}. While 3GPP Releases~15--19 introduced targeted patches---such as Configured Grant (CG) uplink, PUSCH blind repetitions, HARQ feedback disabling for non-terrestrial networks (NTN), and extended FR2-2 numerologies (480/960~kHz)---deployed radio stacks still isolate CE, FEC, CRC, and HARQ behind hard interfaces. Recommendation ITU-R M.2160 for IMT-2030 (6G)~\cite{itu2023} exposes the limits of these incremental patches by expanding the IMT-2020 triangle (eMBB, mMTC, URLLC) into six coupled usage scenarios: Immersive Communication, Hyper Reliable and Low-Latency Communication (HRLLC), Massive Communication, Ubiquitous Connectivity, Integrated Artificial Intelligence (AI) and Communication, and Integrated Sensing and Communication (ISAC).
	
	Beneath these high-level scenarios lies a standardization gap for machine-type communication (MTC). Unlike human-centric multimedia streams characterized by long transport blocks and generous delay budgets, next-generation MTC---spanning industrial closed-loop actuation, massive telemetry, autonomous swarms, and distributed edge inference---operates predominantly with short packets, sporadic bursts, tight coherence blocks, and severe energy or latency ceilings~\cite{durisi2016}. The core challenge is not that academia has overlooked short-packet reliability, but that bridging cross-layer academic solutions into interoperable 3GPP Release~20+ specifications remains unresolved across the classical $\text{CE} \rightarrow \text{FEC} \rightarrow \text{CRC} \rightarrow \text{HARQ}$ chain.

	\begin{figure*}[t]
		\centering
		\IfFileExists{fig_scenarios_bottlenecks.pdf}{%
			\includegraphics[width=\textwidth]{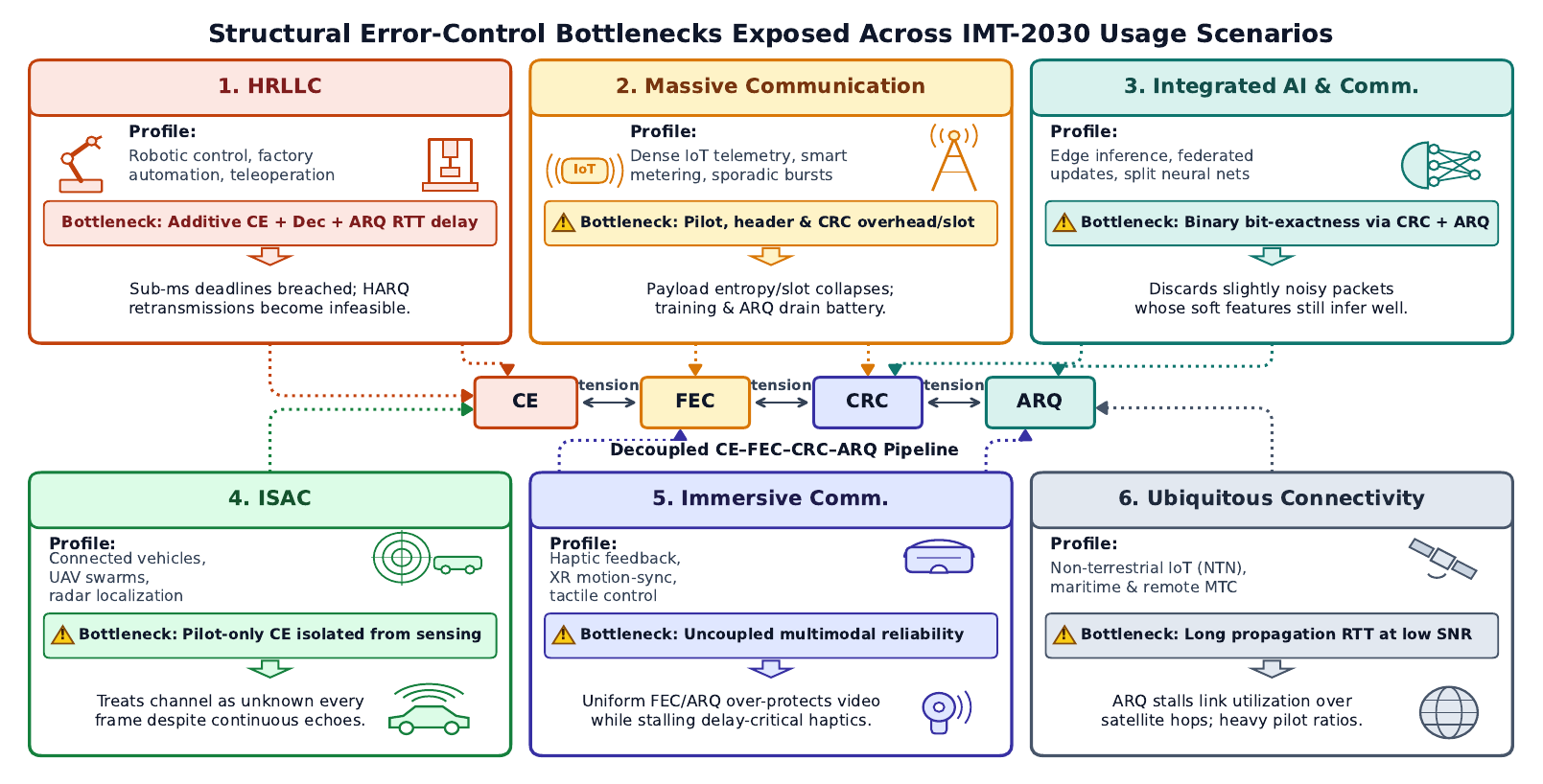}%
		}{%
			\includegraphics[width=0.76\textwidth]{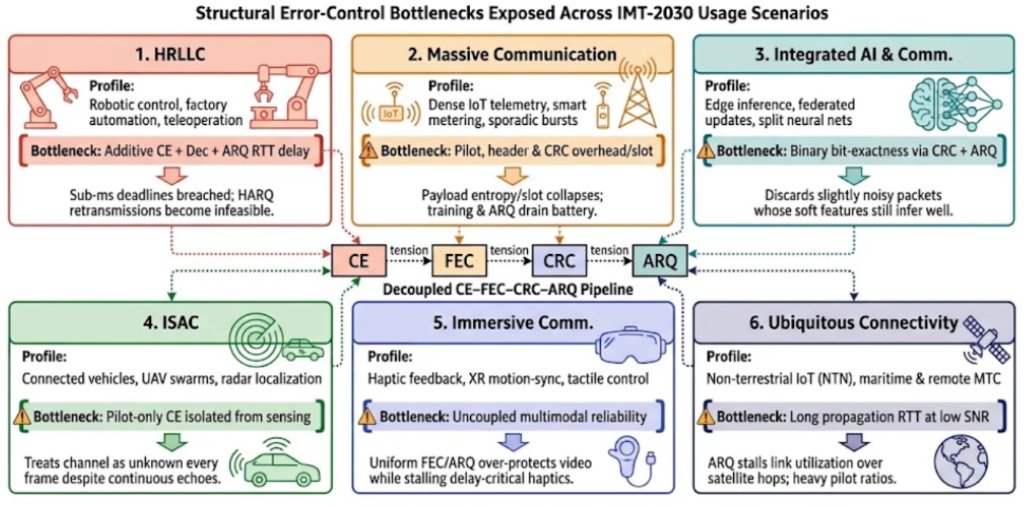}%
		}
		\caption{Structural error-control bottlenecks exposed across the six IMT-2030 usage scenarios, illustrating representative MTC/service profiles, dominant legacy bottlenecks, and the resulting tension with the decoupled CE--FEC--CRC--ARQ pipeline~\cite{popovski2018,durisi2016,senel2018,gunduz2023,liu2022,hassibi2003}.}
		\label{fig:scenarios_bottlenecks}
	\end{figure*}
	
	In this article, we argue that structured fault tolerance requires architectural unification rather than isolated code tuning for IMT-2030, offering three contributions:
	\begin{itemize}
		\item \textbf{Entropy-per-slot characterization:} We examine how allocating finite coherence-block slots to pilot training, FEC parity, and CRC tails degrades net payload entropy resolved per channel use under short-packet MTC constraints.
		\item \textbf{Quantitative bottleneck diagnosis:} We separate true information-loss interfaces (pilot-only CSI truncation and PHY/MAC-to-application CRC collapse) from coupled temporal bottlenecks, substantiating all three via short-packet SO-GRAND simulations and 3GPP TS~38.214 processing timelines~\cite{ts38214}.
		\item \textbf{Standardization-aware IMT-2030 roadmap:} Building beyond 3GPP Releases~15--19, we synthesize four reliability paradigms---joint estimation with soft-output universal decoding~\cite{yuan2025sogrand}, ISAC-assisted priors~\cite{wang2025isac}, grant-free joint detection, and task-oriented error control---alongside their physical trade-offs and Release~20+ hurdles.
	\end{itemize}
	
	\section{The Entropy-per-Slot Dilemma in Short-Packet MTC}
	Classical error-control theory was forged in the asymptotic blocklength regime, where the number of channel uses $N$ per codeword grows arbitrarily large. In that regime, the law of large numbers averages out additive noise fluctuations, channel capacity accurately predicts maximum sustainable throughput, and fixed-size metadata---such as synchronization preambles, training pilots, protocol headers, and 16- or 24-bit CRC tails---represents a negligible fraction of the transmitted frame. Consequently, engineers could treat channel estimation, channel coding, and error detection as largely independent optimization subproblems.
	
	Short-packet MTC shatters this comfortable separation. As formalized by finite-blocklength information theory, when a packet carries only a few dozen to a few hundred information bits, the maximum achievable coding rate backs off sharply from Shannon capacity by a penalty term governed by the channel dispersion and the inverse square root of the blocklength~\cite{polyanskiy2010}. Crucially, the total number of time-frequency resource elements (slots) available inside a single fading coherence block is strictly bounded. Every symbol slot inside that frame must perform one of four competing roles: resolving channel uncertainty (pilots), protecting against noise and fading (FEC redundancy), validating codeword integrity (CRC parity), or conveying actual application information (payload).
	
	From an information-theoretic perspective, adding redundancy bits to protect a short payload creates an intrinsic \textit{entropy-per-slot trade-off}. While FEC parity increases minimum Hamming distance and suppresses block error rates, every slot consumed by redundant bits or pilot symbols lowers the net payload entropy resolved per slot used for communication. In massive MTC deployments, metadata and reliability overhead can easily match or exceed the payload itself~\cite{durisi2016}. For instance, transmitting a 32-bit industrial sensor reading with a 16-bit CRC and a rate-$1/3$ short channel code yields 144 coded bits; under BPSK modulation (1~bit/symbol) and 20~percent pilot overhead, the frame requires 180 channel uses (or 90 channel uses under QPSK)---meaning over 82~percent of the air-interface resource budget is spent on supporting reliability and estimation rather than carrying fresh source entropy. At scale, this structural inflation directly threatens the IMT-2030 sustainability pillar by multiplying radio-on time and battery drain across millions of low-power devices.
	
	\begin{figure*}[t]
		\centering
		\IfFileExists{fig_bottleneck_pipeline.pdf}{%
			\includegraphics[width=\textwidth]{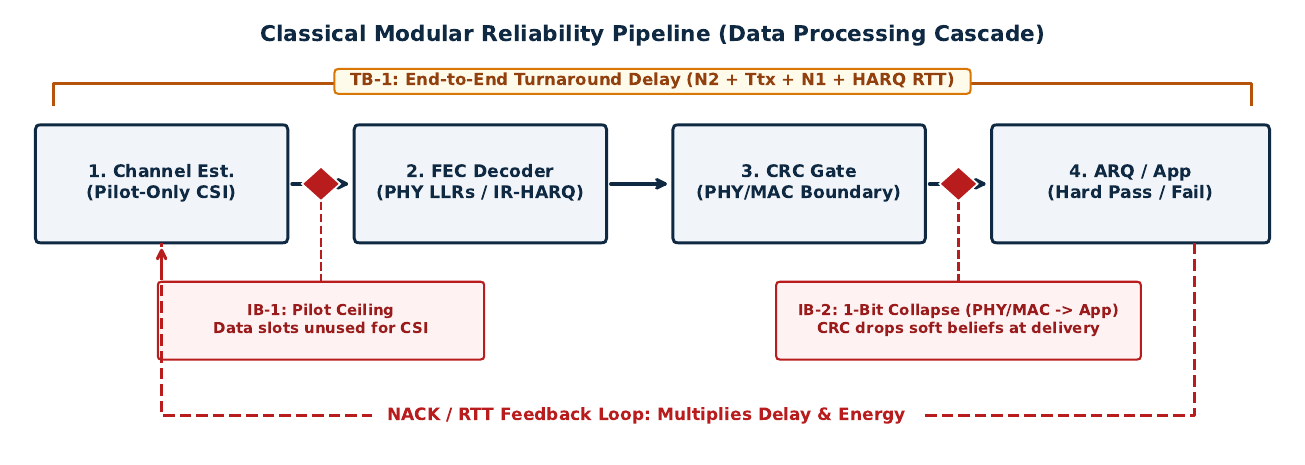}%
		}{%
			\fbox{\parbox[c][4.2cm][c]{0.88\textwidth}{\centering\textbf{[Fig. 3 Placeholder]}\\[4pt]Run \texttt{python figs.py} to generate \texttt{fig\_bottleneck\_pipeline.pdf}.}}%
		}
		\caption{Information bottlenecks (IB-1, IB-2) and temporal bottleneck (TB-1) embedded in the conventional decoupled reliability cascade. Hard layer interfaces restrict mutual-information flow across time-frequency resources and compound standardized processing latency under short-packet MTC.}
		\label{fig:bottlenecks}
	\end{figure*}
	
	Fig.~\ref{fig:scenarios_bottlenecks} maps how each of the six IMT-2030 usage scenarios stresses the classical reliability stack in distinct, mutually reinforcing ways. Hence, optimizing a single parameter inside the legacy pipeline tends to worsen another: strengthening FEC increases decoding complexity and latency; shrinking the pilot allocation to save slots degrades channel estimation accuracy and increases block errors; shortening the CRC reduces overhead but elevates undetected false-alarm rates in list decoding~\cite{shirvanimoghaddam2019}. These tightly coupled failure modes signal a deeper architectural constraint.
	
	\section{Cascaded Information and Temporal Bottlenecks}
	A rigorous way to diagnose the structural fragility of legacy error control is to view the receiver pipeline as a Markov chain of lossy transformations. By the Data Processing Inequality, any sequential stage that compresses or discards statistical information about the transmitted message or the propagation channel can never have that lost mutual information recovered by downstream blocks~\cite{cover2006}. Through this lens, we distinguish two true information bottlenecks (\textbf{IB-1} and \textbf{IB-2})---interfaces that irreversibly restrict the statistical information available to downstream blocks and services---from one coupled temporal bottleneck (\textbf{TB-1}) across the conventional $\text{CE} \rightarrow \text{FEC} \rightarrow \text{CRC} \rightarrow \text{HARQ}$ chain, as summarized in Fig.~\ref{fig:bottlenecks}.
	
	\subsection{IB-1: Pilot Ceiling in Multi-User Short-Packet Channel Estimation}
	The first information bottleneck arises at the boundary between channel estimation and decoding. In conventional receivers, the channel is estimated exclusively from $N_p$ known training pilots placed within a frame of $N$ total symbols~\cite{hassibi2003}, and then frozen as a static input to the equalizer and FEC decoder. While pilot overhead is benign in single-antenna scalar channels at moderate SNR, it becomes acute in \textbf{multi-antenna and multi-user MTC} short frames, where $K$ concurrent devices must share orthogonal pilot sequences inside a tight coherence block~\cite{senel2018}.
	
\begin{figure*}[t]
		\centering
		\IfFileExists{fig_uncertainty_frame.pdf}{%
			\includegraphics[width=\textwidth]{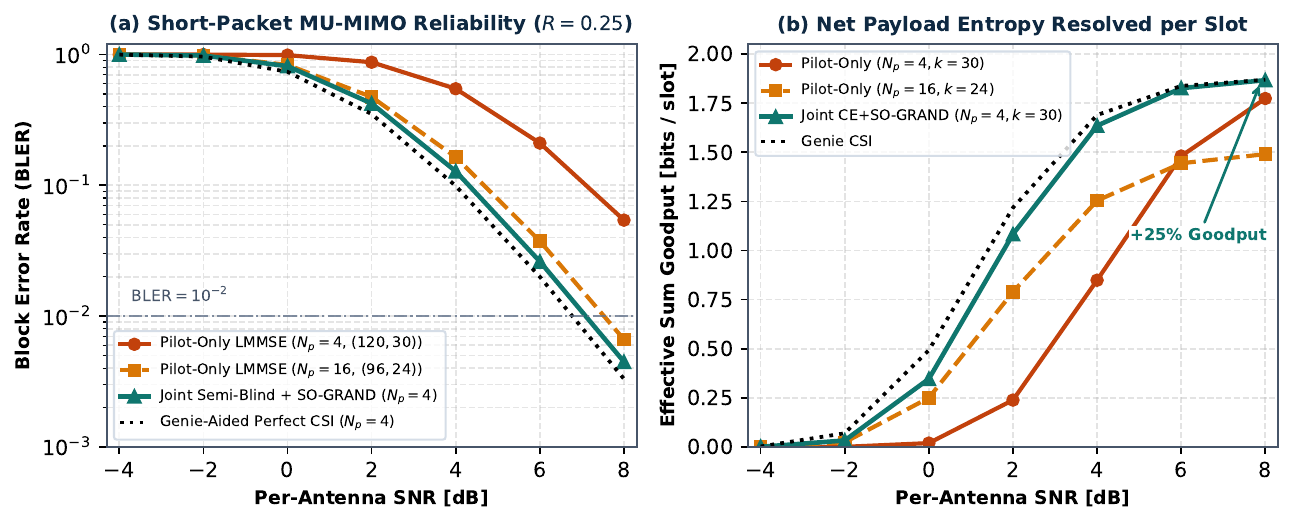}%
		}{%
			\fbox{\parbox[c][5.0cm][c]{0.88\textwidth}{\centering\textbf{[Fig. 3 Placeholder]}\\[4pt]Run \texttt{python figs.py} to generate \texttt{fig\_uncertainty\_frame.pdf}.}}%
		}
		\caption{Link-level Monte Carlo simulation of a short-packet MU-MIMO uplink ($K=4$ UEs, $M=8$ BS antennas, $N=64$ channel uses, QPSK, fixed rate $R=0.25$, single-block SO-GRAND decoding, up to $150{,}000$ frames/point): (a) Block Error Rate (BLER) and (b) effective sum goodput vs. per-antenna SNR. Joint semi-blind CE+SO-GRAND with $N_p=4$ pilots surpasses decoupled pilot-only LMMSE with $N_p=16$ pilots in both reliability and net goodput (+25~percent).}
		\label{fig:uncertainty}
	\end{figure*}

	To quantify this bottleneck concretely, Fig.~\ref{fig:uncertainty} reports a link-level Monte Carlo simulation (scaling from $4{,}000$ frames at low SNR up to $150{,}000$ frames at $+8~\text{dB}$ to resolve $\text{BLER}$ below $10^{-2}$ without sampling floors) of a short-packet multi-user MIMO uplink with $K = 4$ single-antenna MTC devices transmitting concurrently to an $M = 8$ antenna base station over a Rayleigh block-fading frame of $N = 64$ channel uses (QPSK modulation). Each device encodes a single systematic linear block at fixed code rate $R = 0.25$ using Soft-Output Guessing Random Additive Noise Decoding (SO-GRAND)~\cite{yuan2025sogrand} in logistic-weight order (up to $Q_{\max}=1{,}500$ syndrome queries and $L=3$ candidate codewords): $N_p = 4$ pilots (one per device) leaves $N_d = 60$ data slots for a $(120,30)$ code ($k=30$ bits/UE), whereas $N_p = 16$ pilots leaves $N_d = 48$ slots for a $(96,24)$ code ($k=24$ bits/UE). Under decoupled pilot-only LMMSE estimation, $N_p = 4$ suffers severe CSI noise and inter-user interference ($\text{BLER} > 5\times 10^{-2}$ at $+8~\text{dB}$), while $N_p = 16$ reaches $\text{BLER} = 10^{-2}$ at $\sim 7.5~\text{dB}$ at the cost of 25~percent pilot overhead. By contrast, a joint semi-blind receiver that starts with $N_p = 4$ pilots and runs three iterative LMMSE channel updates using SO-GRAND's posterior soft symbol expectations reaches $\text{BLER} = 10^{-2}$ at $\sim 7.0~\text{dB}$---outperforming the $4\times$-pilot-heavy baseline by $0.5~\text{dB}$ while delivering $1.87\text{ bits/slot}$ versus $1.49\text{ bits/slot}$ at $+8~\text{dB}$, a $+25$~percent net goodput gain at identical code rate (Fig.~\ref{fig:uncertainty}b). Although the three semi-blind iterations add gNB baseband compute, each iteration takes orders of magnitude less time than a millisecond-scale HARQ round-trip, yielding a net latency and energy win whenever it prevents a retransmission. Note that iterative semi-blind estimation can exhibit error propagation at very low SNR if initial symbol beliefs are unreliable, motivating low-complexity soft-output initialization.
	
	\subsection{TB-1: Standardized 3GPP Processing Latency and HARQ Turnaround}
\begin{figure*}[t]
		\centering
		\IfFileExists{fig_tb1_ib2.pdf}{%
			\includegraphics[width=\textwidth]{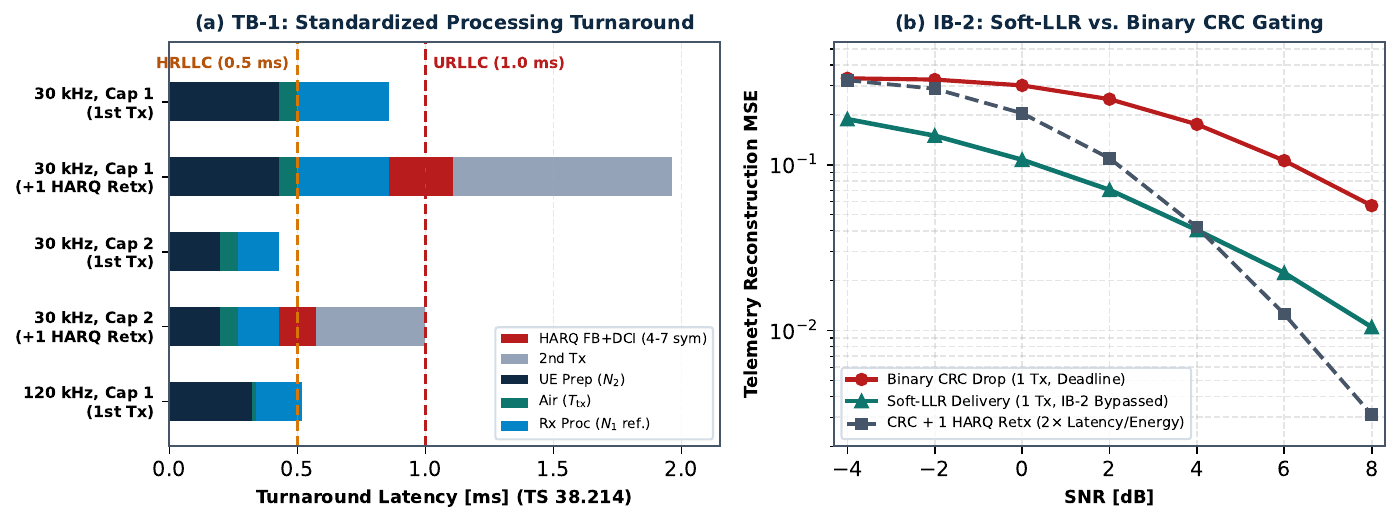}%
		}{%
			\fbox{\parbox[c][4.2cm][c]{0.88\textwidth}{\centering\textbf{[Fig. 4 Placeholder]}\\[4pt]Run \texttt{python figs.py} to generate \texttt{fig\_tb1\_ib2.pdf}.}}%
		}
		\caption{Quantifying temporal and soft-information bottlenecks: (a) \textbf{TB-1} physical-layer turnaround latency for a 2-symbol mini-slot using 3GPP TS~38.214 UE processing times ($N_1, N_2$)~\cite{ts38214} with 4--7 symbols for HARQ feedback/DCI; (b) \textbf{IB-2} telemetry reconstruction MSE ($40{,}000$ Rayleigh 1$\times$2 frames, 32-bit payload + 16-bit CRC, $(144,48)$ SO-GRAND), showing soft-LLR delivery cutting MSE by 43--77\% over binary CRC drops.}
		\label{fig:tb1_ib2}
	\end{figure*}

	The second bottleneck is temporal rather than information-theoretic. In long-packet broadband links, air-interface transmission duration dominates the end-to-end delay budget, masking computational processing times. In IMT-2030 HRLLC and delay-sensitive MTC, however, transmissions shrink to mini-slots of only two OFDM symbols. Under these conditions, standardized 3GPP physical-layer turnaround timelines become the dominant limiter:
	\begin{equation}
		T_{\text{total}} = T_{\text{prep}}(N_2) + T_{\text{tx}} + T_{\text{dec}}(N_1) + N_{\text{retx}}\,T_{\text{HARQ}}.
		\label{eq:latency}
	\end{equation}
	Here, $T_{\text{prep}}(N_2)$ is the standardized UE PUSCH preparation time ($N_2$ symbols) from dynamic scheduling grant to transmission, $T_{\text{tx}}$ is the over-the-air duration, $T_{\text{dec}}(N_1)$ adopts the standardized UE PDSCH-to-ACK processing budget ($N_1$ symbols) as a conservative physical-layer reference for receiver CE, demodulation, FEC decoding, and CRC checking (in uplink MTC, gNB decoding is not bound by $N_1$ but faces comparable baseband turnaround), and $T_{\text{HARQ}}$ adds feedback/DCI turnaround plus a full retransmission cycle. Note that while Configured Grant (CG) uplink eliminates dynamic DCI decoding overhead inside $N_2$, internal UE preparation and gNB decoding still bound turnaround. Using exact symbol counts from 3GPP TS~38.214~\cite{ts38214} and assuming 7 symbols ($0.250~\text{ms}$) under Cap~1 or 4 symbols ($0.143~\text{ms}$) under Cap~2 for HARQ feedback and DCI turnaround without additional slot-alignment wait, Fig.~\ref{fig:tb1_ib2}a shows that at $30~\text{kHz}$ subcarrier spacing ($\mu = 1$, $T_{\text{sym}} = 35.7~\mu\text{s}$), a 2-symbol mini-slot takes only $0.071~\text{ms}$ on air, whereas UE Capability~1 processing ($N_2=12$, $N_1=10$) consumes $0.786~\text{ms}$ ($11\times$ air time), and a single HARQ retransmission pushes latency to $1.96~\text{ms}$---nearly double the $1.0~\text{ms}$ 5G URLLC ceiling. Even with low-latency UE Capability~2 ($N_2=5.5$, $N_1=4.5$), one HARQ cycle hits $1.00~\text{ms}$, breaching sub-millisecond ($0.5~\text{ms}$) HRLLC targets~\cite{popovski2018}. Scaling to $120~\text{kHz}$ ($\mu = 3$, where Capability~2 is unsupported in TS~38.214) shrinks $T_{\text{tx}}$ to $0.018~\text{ms}$ while $N_2+N_1=56$ symbols still consume $0.50~\text{ms}$ ($28\times$ air time); Release~17/19 FR2-2 numerologies (480 and 960~kHz) further compress symbol duration, yet still require tight UE/gNB pipeline integration in Release~20+ study items to avoid processing-dominated stalls.
	
	\subsection{IB-2: Binary Soft-Information Collapse at the PHY/MAC--Application Boundary}
	The second true information bottleneck occurs at the boundary where the CRC verifier mediates delivery to upper layers and applications. Although incremental-redundancy HARQ (IR-HARQ) preserves soft LLRs inside the physical-layer buffer for combining across retransmissions, the CRC stage collapses the high-dimensional posterior bit distribution into a single binary verdict---ACK (1) or NACK (0)---at the PHY/MAC-to-application interface. Even when 98~percent of the message entropy has been resolved with high confidence, a single residual bit flip triggers a binary rejection: under hard deadline constraints the packet is dropped, and under HARQ the link must pay control-channel signaling overhead, feedback latency, and buffer retention cost to coordinate a retransmission.
	
	For emerging IMT-2030 services such as Integrated AI and Communication or massive environmental sensing, this rigid bit-exactness contract is fundamentally misaligned with application utility~\cite{gunduz2023}. To quantify IB-2, Fig.~\ref{fig:tb1_ib2}b simulates the 32-bit telemetry vector from Section~II (four uniform $[-1,1]$ coordinates quantized to 8~bits each) protected by a 16-bit CRC and a rate-$1/3$ $(144,48)$ code over a 1$\times$2 SIMO Rayleigh channel with SO-GRAND. Under a single-transmission deadline, bypassing the binary CRC gate and reconstructing each coordinate from its posterior bit expectations (soft-LLR delivery) reduces reconstruction MSE by 43--77~percent across $-4$ to $+4~\text{dB}$ ($0.189 \rightarrow 0.040$ vs.\ $0.332 \rightarrow 0.175$ for binary CRC drops), matching or outperforming a full Chase-combining HARQ retransmission up to $+4~\text{dB}$ at half the latency and radio-on energy.
	
	\section{Rethinking Reliability Paradigms for IMT-2030 MTC}
\begin{figure*}[t]
		\centering
		\IfFileExists{fig_architecture_comparison.pdf}{%
			\includegraphics[width=\textwidth]{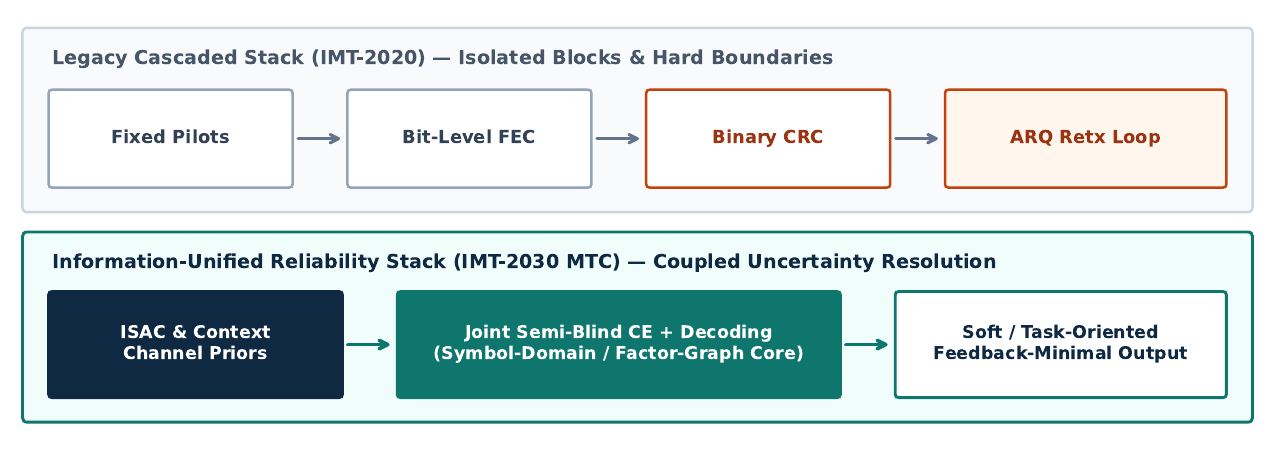}%
		}{%
			\fbox{\parbox[c][4.2cm][c]{0.88\textwidth}{\centering\textbf{[Fig. 5 Placeholder]}\\[4pt]Run \texttt{python figs.py} to generate \texttt{fig\_architecture\_comparison.pdf}.}}%
		}
		\caption{Architectural contrast between the legacy IMT-2020 cascaded stack and an information-unified IMT-2030 reliability architecture designed to eliminate pilot ceilings, additive processing stalls, and binary soft-information collapse in MTC.}
		\label{fig:architecture}
	\end{figure*}

	Overcoming the three bottlenecks identified above requires shifting from isolated block optimization toward \textit{information-unified reliability architectures}. Rather than abandoning structured fault tolerance, next-generation MTC systems must dissolve the rigid interfaces between channel acquisition, error protection, and application-level verification. Fig.~\ref{fig:architecture} contrasts the legacy cascaded IMT-2020 pipeline with an integrated IMT-2030 reliability stack, while Table~\ref{tab:roadmap} synthesizes the engineering trade-offs of its four complementary pillars.
	
	\begin{table*}[h]
		\centering
		\scriptsize
		\caption{Information-Unified Reliability Paradigms for 3GPP Release~20+ IMT-2030 MTC}
		\label{tab:roadmap}
		\renewcommand{\arraystretch}{1.02}
		\begin{tabular}{@{} >{\raggedright\arraybackslash}p{2.9cm} >{\raggedright\arraybackslash}p{1.2cm} >{\raggedright\arraybackslash}p{2.7cm} >{\raggedright\arraybackslash}p{2.2cm} >{\raggedright\arraybackslash}p{3.2cm} >{\raggedright\arraybackslash}p{2.9cm} @{}}
			\toprule
			\textbf{Paradigm} & \textbf{Target} & \textbf{Expected Gain} & \textbf{Compute Split} & \textbf{3GPP / O-RAN Impact} & \textbf{Primary Failure Modes} \\
			\midrule
			\textbf{Joint Semi-Blind CE + SO-GRAND} & IB-1 & $+25\%$ net goodput at $R{=}0.25$; $4\times$ fewer pilots & Heavy gNB Rx; minimal UE change & Iterative LMMSE--decoder loop inside L1 DU & Low-SNR error propagation if soft init fails \\
			\textbf{ISAC-Assisted Channel Priors} & IB-1, TB-1 & Shrinks pilot overhead \& estimator convergence delay & gNB / RAN sensing fusion & Cross-function CSI prior API (sensing $\rightarrow$ L1 CE) & NLoS multipath or cross-band bias (FR3 $\rightarrow$ FR2) \\
			\textbf{Grant-Free Joint Detection \& Rateless} & TB-1, IB-1 & Eliminates DCI/HARQ RTT; avoids pilot exhaustion & Joint sparse + short-code Rx at gNB & Extends Rel-15--19 CG/NTN beyond orthogonal pilots & Activity collisions under bursty overload \\
			\textbf{Task-Oriented Soft Error Control} & IB-2, TB-1 & 43--77\% MSE drop; halves energy/latency vs.\ HARQ & App/inference weights soft LLRs & Standardized confidence metadata across PHY/MAC/App & Undetected integrity risk; adversarial soft inputs \\
			\bottomrule
		\end{tabular}
		\vspace{-0.5em}
	\end{table*}
	
	\subsection{Joint Semi-Blind Channel Estimation and Soft-Output Universal Decoding}
	To overcome the MU-MIMO pilot ceiling shown in Fig.~\ref{fig:uncertainty}, future MTC receivers must treat coded data symbols as dual-purpose carriers of both message entropy and channel information. Through factor-graph message passing or iterative LMMSE refinement, soft symbol beliefs produced during decoding refine channel estimates without dense pilot preambles. Recent advances in noise-centric universal decoding---notably Soft-Output Guessing Random Additive Noise Decoding (SOGRAND)~\cite{yuan2025sogrand}---generate accurate \textit{a posteriori} codeword probabilities and explicit correctness confidence for arbitrary short linear codes, outperforming 5G LDPC baselines at short blocklengths while naturally feeding soft symbol priors back to the channel estimator. In ultra-short bursts, non-coherent Grassmannian or compressive-sensing-assisted detectors bypass explicit CSI altogether~\cite{senel2018}, though designers must guard against iterative error propagation at very low SNR.
	
	\subsection{Sensing-Assisted Channel Priors via ISAC}
	One of IMT-2030's defining architectural opportunities lies in exploiting ISAC as an internal enabler for physical-layer error control~\cite{liu2022,wang2025isac}. When the radio access network tracks scatterer geometry, device trajectories, and Doppler shifts through integrated sensing echoes, that environmental state supplies a tight statistical prior for channel estimation, shrinking initial channel uncertainty and reducing both pilot slots and estimator convergence latency. Crucially, this coupling is not universally free: geometric sensing priors correlate strongly with communication channels under line-of-sight (LoS) or dominant-scatterer propagation in co-located frequency bands, whereas rich non-LoS (NLoS) multipath or cross-band operation (e.g., FR3 sensing aiding FR2 links) can weaken correlation or bias LMMSE priors unless paired with robust tolerance factors~\cite{wang2025isac}.
	
	\subsection{Beyond 3GPP Rel-15--19: Joint Grant-Free and Rateless Reliability}
	Where HRLLC deadlines or non-terrestrial propagation delays render multi-round HARQ unviable, 3GPP Releases~15--19 already standardized key open-loop building blocks: Configured Grant (Type~1/2) uplink without dynamic DCI handshakes, mini-slot PUSCH blind repetitions (Type~B), PDCP duplication, HARQ feedback disabling for NTN, and extended FR2-2 numerologies~\cite{popovski2018}. However, these mechanisms still rely on rigid RRC pre-configuration and orthogonal pilot assignment. Extending beyond Rel-19 for IMT-2030 requires \textit{joint} grant-free activity, CSI, and data detection~\cite{senel2018} paired with low-latency short codes (e.g., BCH or SOGRAND-decoded product codes~\cite{shirvanimoghaddam2019,yuan2025sogrand}) and rateless mutual-information accumulation, avoiding both pilot exhaustion and brittle code-rate mismatches.
	
	\subsection{Soft-Information Preservation and Task-Oriented Error Control}
	Finally, IMT-2030 must relax the dogma that every packet requires bit-exact recovery enforced by a rigid binary CRC gate. In Integrated AI and Communication workflows---such as split edge inference, remote state estimation, and anomaly detection---reliability should be governed by \textit{task distortion} rather than raw bit error rate~\cite{gunduz2023}. Passing calibrated soft reliability metrics---such as SOGRAND's \textit{a posteriori} block-correctness probabilities~\cite{yuan2025sogrand}---or unequal-error-protected semantic features across the link layer enables downstream inference engines to weight uncertain features gracefully, suppressing unnecessary HARQ retransmissions whenever task confidence exceeds operational thresholds. Legacy bit-exact CRC--HARQ remains indispensable for firmware updates, cryptographic key exchange, and discrete actuation commands; the architectural shift lies in making reliability contracts service-aware.
	
	\section{Open Research Challenges and Standardization Outlook}
	Translating an information-unified reliability paradigm into deployable IMT-2030 specifications entails nontrivial engineering and standardization hurdles:
	\begin{itemize}
		\item \textbf{Complexity Asymmetry and Hardware Constraints:} Joint channel estimation and decoding, factor-graph solvers, and ordered-statistics decoders shift substantial computational weight to the receiver. While this asymmetry fits uplink MTC---where cloud-native base stations can absorb processing complexity to spare battery-powered sensors---downlink actuation and device-to-device sidelinks require ultra-low-power, hardware-friendly approximations with bounded latency.
		\item \textbf{Cross-Layer Standardization Friction:} Existing 3GPP radio specifications enforce clean functional splits between physical-layer reference signals, channel coding chains, and medium-access-control (MAC) HARQ entities. Introducing semi-blind estimation, ISAC-informed CSI priors, or soft-output link interfaces---such as passing calibrated block-confidence metadata or quantized LLR summaries across PHY--MAC and Open RAN fronthaul/F1 boundaries---will demand carefully designed reference interfaces during upcoming 6G study items (3GPP Release 20 and beyond) to preserve multi-vendor interoperability and Open RAN modularity.
		\item \textbf{Information-Theoretic Benchmarking under Finite Latency:} Classical metrics such as block error rate versus SNR fail to capture the true cost of reliability when pilot overhead, decoder latency, and task accuracy interact. New standardized evaluation methodologies must jointly report net payload entropy efficiency per coherence block, tail latency including $T_{\text{enc}} + T_{\text{CE}} + T_{\text{dec}}$, and energy per successfully completed machine task.
		\item \textbf{Security and Resilience of Reduced-Redundancy Links:} Trimming explicit CRC length or exposing soft beliefs across protocol layers introduces subtle integrity risks, including higher undetected error probabilities and adversarial perturbations against task-oriented receivers. Co-designing lightweight cryptographic authentication with physical-layer error control will be essential to uphold IMT-2030's security and resilience mandate~\cite{itu2023}.
	\end{itemize}
	
	\section{Conclusion}
	\enlargethispage{0.25in}
	Structured fault tolerance and error control in mobile networks are not dying disciplines; rather, they are reaching the boundaries of an architectural template built for asymptotic broadband traffic. Under the finite-blocklength, ultra-low-latency, and energy-constrained realities of IMT-2030 machine-type services, the classical cascade of pilot-only channel estimation, bit-level FEC, binary CRC, and feedback-driven ARQ degrades net payload entropy per slot, inflates turnaround latency, and discards soft beliefs at rigid layer interfaces.
	
	Addressing these bottlenecks calls for a deliberate shift toward information-unified reliability design. By coupling channel acquisition with universal soft-output decoding, harvesting geometric priors from ISAC, minimizing reactive feedback loops, and aligning error verification with application-level task utility, the wireless community can unlock fault-tolerant protocols tailored to the full breadth of IMT-2030 services.
	
	\section*{Acknowledgment}
	The author used Google Gemini for language editing and simulation/plotting code assistance. All technical claims, parameters, numerical results, and conclusions were independently verified by the author, who assumes full responsibility.
	\vspace{-20px}
	\renewcommand{\IEEEbibitemsep}{0pt plus 0.5pt}
	\bibliographystyle{IEEEtran}
	\bibliography{references}

@misc{itu2023,
  author       = {{ITU-R}},
  title        = {Framework and Overall Objectives of the Future Development of {IMT} for 2030 and Beyond},
  howpublished = {Recommendation ITU-R M.2160-0},
  month        = nov,
  year         = {2023}
}

@article{durisi2016,
  author  = {G. Durisi and T. Koch and P. Popovski},
  journal = {Proceedings of the IEEE},
  title   = {Toward Massive, Ultra-Reliable, and Low-Latency Wireless Communication With Short Packets},
  year    = {2016},
  volume  = {104},
  number  = {9},
  pages   = {1711--1726},
  month   = sep
}

@article{polyanskiy2010,
  author  = {Y. Polyanskiy and H. V. Poor and S. Verd{\'u}},
  journal = {IEEE Transactions on Information Theory},
  title   = {Channel Coding Rate in the Finite Blocklength Regime},
  year    = {2010},
  volume  = {56},
  number  = {5},
  pages   = {2307--2359},
  month   = may
}

@article{popovski2018,
  author  = {P. Popovski and J. J. Nielsen and C. Stefanovic and E. de Carvalho and E. Str{\"o}m and K. F. Trillingsgaard and A.-S. Bana and D. M. Kim and R. Kotaba and J. Park and R. B. S{\o}rensen},
  journal = {IEEE Network},
  title   = {Wireless Access for Ultra-Reliable Low-Latency Communication: Principles and Building Blocks},
  year    = {2018},
  volume  = {32},
  number  = {2},
  pages   = {16--23},
  month   = mar
}

@article{senel2018,
  author  = {K. Senel and E. G. Larsson},
  journal = {IEEE Transactions on Communications},
  title   = {Grant-Free Massive {MTC}-Enabled Massive {MIMO}: A Compressive Sensing Approach},
  year    = {2018},
  volume  = {66},
  number  = {12},
  pages   = {6164--6175},
  month   = dec
}

@article{gunduz2023,
  author  = {D. G{\"u}nd{\"u}z and Z. Qin and I. E. Aguerri and H. S. Dhillon and Z. Yang and A. Yener and K. K. Wong and C.-B. Chae},
  journal = {IEEE Journal on Selected Areas in Communications},
  title   = {Beyond Transmitting Bits: Context, Semantics, and Task-Oriented Communications},
  year    = {2023},
  volume  = {41},
  number  = {1},
  pages   = {5--41},
  month   = jan
}

@article{liu2022,
  author  = {F. Liu and Y. Cui and C. Masouros and J. Xu and T. X. Han and Y. C. Eldar and S. Buzzi},
  journal = {IEEE Journal on Selected Areas in Communications},
  title   = {Integrated Sensing and Communications: Toward Dual-Functional Wireless Networks for {6G} and Beyond},
  year    = {2022},
  volume  = {40},
  number  = {6},
  pages   = {1728--1767},
  month   = jun
}

@article{hassibi2003,
  author  = {B. Hassibi and B. M. Hochwald},
  journal = {IEEE Transactions on Information Theory},
  title   = {How Much Training is Needed in Multiple-Antenna Wireless Links?},
  year    = {2003},
  volume  = {49},
  number  = {4},
  pages   = {951--963},
  month   = apr
}

@article{shirvanimoghaddam2019,
  author  = {M. Shirvanimoghaddam and M. S. Mohammadi and R. Abbas and A. Minja and C. Yue and B. Matuz and G. Han and Z. Lin and W. Liu and Y. Li and S. Johnson and B. Vucetic},
  journal = {IEEE Communications Magazine},
  title   = {Short Block-Length Codes for Ultra-Reliable Low Latency Communications},
  year    = {2019},
  volume  = {57},
  number  = {2},
  pages   = {130--137},
  month   = feb
}

@book{cover2006,
  author    = {T. M. Cover and J. A. Thomas},
  title     = {Elements of Information Theory},
  edition   = {2nd},
  publisher = {Wiley-Interscience},
  address   = {Hoboken, NJ, USA},
  year      = {2006}
}

@techreport{ts38214,
  author      = {{3GPP}},
  title       = {{NR}; Physical Layer Procedures for Data (Release 18)},
  institution = {3rd Generation Partnership Project (3GPP)},
  type        = {TS},
  number      = {38.214},
  year        = {2024}
}

@article{yuan2025sogrand,
  author  = {P. Yuan and M. M{\'e}dard and K. Galligan and K. R. Duffy},
  journal = {IEEE Transactions on Wireless Communications},
  title   = {Soft-Output ({SO}) {GRAND} and Iterative Decoding to Outperform {LDPC} Codes},
  year    = {2025},
  volume  = {24},
  number  = {4},
  pages   = {3386--3399},
  month   = apr
}

@inproceedings{wang2025isac,
  author    = {S. Wang and A. Tang and X. Wang and W. Qu},
  booktitle = {Proceedings of the IEEE International Conference on Communications (ICC)},
  title     = {Sensing-Assisted Channel Estimation for Bistatic {OFDM} {ISAC} Systems: Framework, Algorithm, and Analysis},
  year      = {2025}
}
	
	\begin{IEEEbiographynophoto}{Igor B. Palhano}
		(Student Member, IEEE) holds a double degree in Telecommunications Engineering from the Federal University of Ceará, Brazil, and the \textit{École Nationale Supérieure d'Ingénieurs de Caen (ENSICAEN)}, France. His research interests include ISAC, error-correction theory, and machine learning for signal processing.
	\end{IEEEbiographynophoto}
	
\end{document}